\documentclass[12pt]{article}
\usepackage[centertags]{amsmath}

\begin{document}
\title{Comment on the ``Invariance of the Tunneling Method for Dynamical Black
  Holes''}
\date{\today}
\author{M. Pizzi\\Physics Department, University of Rome ``La Sapienza'', \\P.le A. Moro, Italy 00185, \\E-mail: pizzi@icra.it}

\maketitle
\begin{abstract}
In the recent paper arXiv:0906.1725 it has been guessed that a
``correct'' treatment of the semi-classical tunneling based on the Hamilton-Jacobi equation gives the standard Hawking formula, contrary to was
asserted in my previous paper arXiv:0904.4572. Here I confirm my calculation showing where the article of Di Criscienzo et al. is wrong.
\end{abstract}

\section{Introduction}

It has now become clear that the time factor has an important role
in the WKB approximation for the tunneling in curved space-times.
Nevertheless many incorrect calculations have been published and are yet being published; in this paper
I will focus on the recently appeared article Ref.\cite{DHNVZ} which goes in the
opposite direction with respect to my previous paper Ref.\cite{Pizzi}.

The authors of Ref.[1] considered the case of a massless particle
moving in the two-dimensional ($r,t$)-section of a spherically symmetric
spacetime. For such case the action
(eikonal) along the classical light-like ray is identically null (or, if you prefer, identically a constant), then no change in the
action can arise in the course of the particle propagation. Therefore all their
results, namely
Eqns.(II.6), (II.14), (II.25), (II.33), (IV.7), (IV.12), and (V.72), are
wrong \textit{a priori} because also no imaginary part in the action can
appear during the motion. No matter Kodama vectors, Hayward gravity surfaces, non-stationary space-times, and dynamical
trapping horizons. All these notions are irrelevant to the question under
interest.

 The impossibility of arising an imaginary part of the action in the
way proposed in [1] is evident also from the following points (which are
valid in relation not only to the Ref.[1], but also to the most of the papers on this
topic):

1. All calculations are done in an infinitesimally-small
neighbourhood of the horizon.

2. An infinitesimally-small neighbourhood of the horizon is homeomorphic to
the flat spacetime, and can be covered by Minkowski coordinates.
Consequently in this neighbourhood, in Minkowski coordinates, the appearing in the
action of an imaginary part is a miracle that cannot happen.

3. The action of a test particle (with or without mass) is a relativistic
invariant. Consequently such imaginary part cannot arise in \emph{any} other
coordinate system.

\section{The EFB gauge (first sub-section of the the Ref.[1]'s section II)}

Let us analyze in more detail the mistake made in Eqns.(II.3)-(II.6) of the section
II of Ref.[1] since all the rest
of the paper is based on similar errors. 

The interval for a spherically symmetric metric in
Eddington-Finkelstein-Bardeen advanced coordinates (Eqn.(II.3) in Ref.[1]) is:%
\begin{equation}
ds^{2}=-e^{2\Psi (v,r)}C(v,r)dv^{2}+2e^{\Psi (v,r)}dvdr+r^{2}d\Omega ^{2}. 
\label{(1)}
\end{equation}
The horizon is located at the points where $C=0.$ The
Hamilton-Jacobi equation for the action $I(v,r)$ for the outgoing mode is:
\begin{equation}
C\partial _{r}I=-2e^{-\Psi }\partial _{v}I.  \label{(2)}
\end{equation}
 This is the Eqn.(II.5) of Ref.[1] after the correction of the sign in its
r.h.s. Indeed the sign in the r.h.s. of Ref.[1]'s Eqn.(II.5) should be minus and
not plus; however, although important, this is not yet the basic mistake. The
action along an outgoing trajectory $v=v(r)$, taking into account Eqn.(2), results from the following evident formulas:%
\begin{equation}
I=\int (\partial _{r}Idr+ \partial _{v}Idv)=\int (\partial _{r}I-\frac{1}{2}%
\frac{dv}{dr}e^{\Psi }C\partial _{r}I)dr.  \label{(3)}
\end{equation}

 To calculate the second term (the ``temporal" one) in the integrand of (3) we
need to know $dv/dr$, which follows from the equation $ds^{2}=0$ for the
outgoing ray:%
\begin{equation}
\frac{dv}{dr}=2e^{-\Psi }C^{-1}  \label{(4)}
\end{equation}

 The substitution of (4) into (3) gives for the integrand in (3)
identically zero. This is a demonstration of the well known fact that we already
mentioned in the introduction, i.e. that the eikonal does not change along the classical ray and
and therefore that no imaginary part can arise in it.

The mistake of the authors of Ref.[1] is in the incorrect derivation of the equation of motion near the horizon. Instead of our formula (4), they wrote
the incorrect equation $dvdr=0$ (see their Eqn. [II.4]), from where they conclude, also with a pure and simple \textsl{non sequitur}, that \textquotedblleft the temporal contribution does not contribute
to the imaginary part of the action\textquotedblright . By other words
their assertion was that the term $\int \partial _{v}Idv$ in the action could be neglected. Then they calculated (see [II.6] of Ref.[1]) the
action taking into account only the first term $\int\partial _{r}Idr$, and using the fact that this integral has a pole on the horizon.

 I am sorry to say that these manipulations are simply wrong since also the temporal term, $\partial _{v}Idv$, 
has on the horizon \emph{the same pole but opposite in sign}, as it can be seen from the formulas
(3)-(4). In the sum the two poles compensate each other and the total integrand has no pole (being just zero), as it should
be expected from the general considerations made in the Introduction. 

\section{Final remarks}

I showed in detail the correct treatment of the tunneling in Eddington-Finkelstein-Bardeen coordinates, showing how the temporal factor has been missed in Ref.[1].

In Ref.[1]'s subsequent sections it has been considered the tunneling in different metrics and coordinates. In these other gauge the authors indeed take into account the temporal factor: but always with the wrong sign, with the consequence that the two poles (the one of the spatial part and the one of the temporal factor) were summed with the same sign instead of being canceled each other.
Unfortunately this sort of errors is characteristic also of many other papers where creditable authors were led to mistake trying to retrieve the believed-correct result of a tunneling.

The exact proof of the impossibility of the tunneling through the horizon
of a black hole has been provided by Belinski \cite{Bel06} and the present Comment
as well as my previous paper \cite{Pizzi} are in agreement with his results. It should be mentioned also that the fact of nonexistence of any poles in the
action's integrand on the horizon due to the contribution from the
temporal term was first mentioned by Belinski in his lectures delivered in
2006 at XII Brazilian School of Cosmology and Gravitation \cite{Bel07}.

\end{document}